\documentclass[12pt,fullpage,doublespace]{article}
\usepackage[english]{babel}
\usepackage[T1]{fontenc}
\usepackage{setspace}
\usepackage{graphicx}
\usepackage{epsfig}
\usepackage[paper=a4paper,left=30mm,right=30mm,top=25mm,bottom=40mm]{geometry}
\usepackage{lscape}
\date{ }
\title{The excitation of solar-like oscillations in a $\delta$ Sct star by efficient envelope convection }
\begin{document}
\maketitle

\author{V. Antoci$^{1\ast} $, G. Handler$^{1,2}$, T.L. Campante $^{3,4}$, A. O. Thygesen$^{4,5}$, A. Moya$^{6}$, T. Kallinger$^{1,7,8}$, D. Stello$^{9}$,  A. Grigahc\`ene$^{3}$, H. Kjeldsen$^{4}$, T.R. Bedding$^{9}$,  T. L\"uftinger$^{1}$, J. Christensen-Dalsgaard$^{4}$, G. Catanzaro$^{10}$, A. Frasca$^{10}$, P. De Cat$^{11}$ , K. Uytterhoeven$^{12,13,14}$, H. Bruntt$^{4}$, G. Houdek$^{1}$, D.W. Kurtz$^{15}$, P. Lenz$^{2}$, A. Kaiser$^{1}$, J. Van Cleve$^{16}$, C.  Allen$^{17}$, B.D. Clarke$^{16}$}

$^{1}$Institute of Astronomy, University of Vienna, T\"urkenschanzstr. 18, A-1180 Vienna, Austria\\
$^{2}$Copernicus Astronomical Center, Bartycka 18, 00-716 Warsaw, Poland \\ 
$^{3}$Centro de Astrof\'isica, DFA-Faculdade de Ci\^encias, Universidade do Porto, Rua das Estrelas, 4150-762 Porto, Portugal\\ 
$^{4}$Department of Physics and Astronomy, Aarhus University, bygn. 1520, Ny Munkegade, DK-8000 Aarhus C, Denmark\\
$^{5}$Nordic Optical Telescope, Apartado 474, E-38700 Santa Cruz de La Palma, Santa Cruz de Tenerife, Spain\\
$^{6}$Departamento de Astrof\'isica. Centro de Astrobiolog\'ia, INTA-CSIC, PO BOX 78, E-28691, Villanueva de la Ca$\tilde{n}$ada, Madrid, Spain \\
$^{7}$ Department of Physics and Astronomy, University of British Columbia, 6224 Agricultural Road, Vancouver, BC V6T 1Z1, Canada\\
$^{8}$Instituut voor Sterrenkunde, K. U. Leuven, Celestijnenlaan 200D, 3001 Leuven, Belgium\\
$^{9}$Sydney Institute for Astronomy (SIfA), School of Physics, University of Sydney, NSW 2006, Australia\\
$^{10}$INAF -- Osservatorio Astrofisico di Catania, via S. Sofia 78, I-95123, Catania, Italy\\
$^{11}$Royal Observatory of Belgium, Ringlaan 3, B-1180 Brussels, Belgium\\
$^{12}$Laboratoire AIM, CEA/DSM-CNRS-Universit\'e Paris Diderot; CEA, IRFU, SAp, Centre de Saclay, F-91191, Gif-sur-Yvette, France\\
$^{13}$Kiepenheuer-Institut f\"ur Sonnenphysik, Sch\"oneckstr. 6, 79104 Freiburg, Germany\\
$^{14}$Instituto de Astrof\'isica de Canarias, 38200 La Laguna, Tenerife, Spain; Departamento de Astrof\'isica, Universidad de La Laguna, 38205 La Laguna, Tenerife, Spain\\
$^{15}$Jeremiah Horrocks Institute, University of Central Lancashire, Preston PR1\,0AQ, UK\\	
$^{16}$SETI Institute/NASA Ames Research Center, Moffett Field, CA 94035, USA\\
$^{17}$Orbital Sciences Corporation/NASA Ames Research Center, Moffett Field, CA 94035, USA\par

\newpage

{\bf 

Delta Scuti ($\delta$ Sct)$^{1}$ stars are opacity-driven pulsators with masses
of 1.5-2.5M$_{\odot}$, their pulsations resulting from the varying ionization of helium. In less massive stars$^{2}$ such as the Sun, convection
transports mass and energy through the outer 30 per cent of the
star and excites a rich spectrum of resonant acoustic modes. Based
on the solar example, with no firm theoretical basis, models predict
that the convective envelope in  $\delta$ Sct stars extends only about 1 per
cent of the radius$^{3}$, but with sufficient energy to excite solar-like
oscillations$^{4,5}$. This was not observed before the Kepler mission$^{6}$, so
the presence of a convective envelope in the models has been
questioned. Here we report the detection of solar-like oscillations
in the $\delta$ Sct star HD 187547, implying that surface convection
operates efficiently in stars about twice as massive as the Sun, as
the ad hoc models predicted.

}

 Thirty days of continuous observations of HD 187547 (KIC 7548479) by the {\it Kepler} mission with a cadence of one minute led to its identification as a $\delta$ Sct pulsator (Fig.~1a, b). In contrast to the non-uniformly distributed signals at low frequencies the observed regularly spaced peaks at high frequencies (Fig.~1c) suggest that we also observe high-radial overtones as expected for stochastically excited solar-like oscillations. For such oscillations the observed comb-like frequency structure (with the large frequency separation $\Delta \nu$ indicating   the frequency separation between consecutive radial overtones of like degree) is the result of mainly radial and dipolar pulsation modes, whereas for $\delta$ Sct stars it is not clear which modes are excited to observable amplitudes. The strikingly broadened structures observed only at high frequencies (Fig.~1f, Fig.~2b, c) suggest that each are either due to single damped and stochastically re-excited oscillations, or very close unresolved frequencies of coherent oscillations.

 Here we use spectroscopic observations to derive an effective temperature ${\mathit{T}_{\rm eff}}$ = 7,500 $\pm$ 250~K, a surface gravity of $\log g$ = 3.90 $\pm$ 0.25 dex [cgs], and a projected rotational velocity of $v \sin i$ = 10.3 $\pm$ 2.3 km~s$^{-1}$ (see Supplementary Information for details). We identify
HD 187547 as an Am star from chemical element abundance analysis,
which is consistent with the observed low $v \sin i$ typical for these stars.
Am stars are stars of spectral type A showing atmospheric underabundance when compared with the Sun in the chemical elements
Sc and Ca, and an overabundance of Ba, Sr and Y (ref. 7). We compute
a photospheric metallicity (all elements except H and He) of Z = 0.017,
which is larger than the solar value of Z = 0.0134 (ref. 8).

 About two-thirds of Am stars are primary components of spectroscopic binary systems$^{9}$. The Am phenomenon is connected to slow rotation, which is not common in A type stars. Binarity is believed to
act as a braking mechanism slowing down the rotation and allowing
spectral peculiarities to occur as a result of element diffusion$^{10}$.
Pulsating Am stars still represent a challenge to theory, because He
is expected to settle gravitationally and should only partly be present in
the He II ionization zone where the $\delta$ Sct pulsations are excited. In
other words, theoretical models predict that the hottest and youngest
A-type stars should not pulsate$^{10}$, which is in contradiction with recent
observations$^{11}$. As the stars evolve, their convective envelopes deepen
and efficiently mix the stellar matter, erasing the observed chemical
peculiarities in the atmospheres, allowing the opacity mechanism to
drive pulsation in the He II ionization zone. Using the observed solar-
like oscillations reported here, the depth of the convective envelope can
be derived (hence the mixing length), probing the diffusion of He and
heavy elements in this star. This will contribute significantly to revising
the interaction between pulsation and diffusion in models of Am stars.

Seven radial velocity measurements of HD 187547, spread over 153 days, give no evidence for a short-period binary system. In addition, the absence of any detectable contribution by a potential close companion to the spectrum implies a considerably less luminous star of spectral type G or later. The expected amplitudes and frequency of maximum oscillation power for such a star are inconsistent with the  observations, leading to the conclusion that the signal observed in Fig. 1c cannot originate from a companion. The observed amplitude spectrum of HD 187547 is not affected by a background star  because the fraction of light in the aperture from neighbouring stars is only 1.5\%. Other chemically peculiar pulsating stars situated, as the $\delta$ Sct
stars, in the classical instability strip in the Hertzsprung-Russell diagram$^{12}$ are the rapidly oscillating Ap stars. Their high-radial-order
pulsation modes are triggered by the opacity mechanism acting in
the hydrogen ionization zone, often showing equidistant multiplets
in the frequency spectrum as a result of the alignment of the pulsation
axes with strong magnetic fields$^{13}$. The strong magnetic fields as seen in
rapidly oscillating Ap stars are, however, not observed in Am stars$^{14}$.
We therefore exclude the possibility that HD 187547 is a hybrid of a
$\delta$ Sct and a rapidly oscillating Ap star.\par

In Fig. 3 we show an \'{e}chelle diagram comparing the observed frequencies with a model of a star similar to HD 187547, demonstrating again the clear structures separated by $\Delta \nu$ at high frequencies and the non-structured distribution at lower frequencies. For the high-frequency modes we derive a mean large frequency separation $\Delta \nu$ of 40.5 $\pm$ 0.6~$\mu$Hz. Using the empirical relation$^{15}$ $\Delta \nu = (0.263\pm 0.009)\mu {\rm Hz}(\nu_{\rm max}/\mu {\rm Hz})^{0.772\pm0.005}$  we obtain a frequency of maximum power $\nu_ {\rm max}$ = 682$^{+41}_{-43}$~$\mu$Hz. This is in very good agreement with the highest amplitude mode in the supposed stochastic frequency region at 696~$\mu$Hz. The possibility that what we observe is 0.5$\Delta \nu$ in the frequency spectrum is ruled out because this would require a $\nu_{\rm max}$ at about  1,673 $\mu$Hz, where no signal is observed. We can also exclude to observe 2$\Delta \nu$ because that would place $\nu_{\rm max}$ at 277 $\mu$Hz, close to the dominant opacity-driven mode at 251 $\mu$Hz.

The amplitudes of solar-like oscillations are determined by the interaction between driving and damping defined by different physical processes$^{2}$, such as modulation of the turbulent momentum and heat fluxes by pulsation. The exact contribution to driving and damping by each of these processes is still not well understood, resulting in uncertainties in the predictions of the stochastically excited mode amplitudes$^{16}$, particularly in hotter stars$^{2,3}$ in which the convective envelopes are shallow.  
 We expect the mixing length, the amplitudes and mode lifetimes to constrain the anisotropy of the convective velocity field, parameters that all
semi-analytical convection models rely on$^{17}$.

For HD~187547 we measure a peak-amplitude per radial mode$^{18}$ for the assumed stochastic signal of 56 $\pm$ 2 p.p.m., which after bolometric correction$^{19}$ results in 67$\pm$ 3 p.p.m. (see Supplementary Information for details). From the empirical scaling relation$^{20}$ and using a bolometric solar peak-amplitude of 3.6 p.p.m.(ref. 21) we obtain a predicted peak-amplitude of A = 14 $\pm$ 9 p.p.m.. The mean mode lifetimes are measured$^{22}$ as 5.7$\pm$0.8 days. Empirical relations predict a mode lifetime for a star with ${\mathit{T}_{\rm eff}}$ = 7,500  $\pm$ 250K of the order of one day$^{23}$ or shorter$^{24}$, which is not in agreement with what we measure for HD~187547. However these scaling relations (for amplitude and mode lifetimes) are based on few observed stars and none of them is calibrated in the temperature domain of our target, for which the physical conditions in the convection zone are expected to be very different. Furthermore, given that HD 187547 is metal overabundant in comparison with the Sun, the observed amplitude is expected to be higher$^{3,25}$ than predicted from simple scaling which is indeed the case. The power of a mode is directly proportional to the mode lifetime provided the energy supply rate over the mode inertia is constant$^{26}$, which further supports the higher amplitude as the observed mode lifetimes are also longer than expected. An additional factor which is not considered in any scaling relations is the chemical peculiarity of our target. In summary, these factors make HD~187547 an intriguing case for further theoretical analyses of stochastic oscillations and the potential interaction with the opacity mechanism in $\delta$ Sct stars.

The amplitude distribution for stochastic pulsation can be described as a Rayleigh distribution, provided the examined time series are much shorter than the mode lifetimes. The relation between the mean amplitude $\langle A \rangle$ and its standard deviation $\sigma(A)$ can then be written as$^{27}$ $(4/\pi - 1)^{0.5}\langle A \rangle \simeq 0.52 \langle A \rangle$. This is not valid for opacity-driven pulsation. In the case of HD 187547 we therefore expect to obtain two different regimes of the ratio $\sigma(A)$/$\langle A \rangle$ for the two groups of oscillation modes (Supplementary Fig. 2). Indeed, we see that the $\delta$ Sct frequencies have a lower value of $\sigma(A)$/$\langle A \rangle$ than the supposed solar-like modes, giving further evidence for the stochastic nature of the latter (see  Supplementary Information for details).

We cannot strictly rule out that the signals between 578 and 868 $\mu$Hz are due to unresolved modes of pulsation excited by the opacity mechanism, because high-radial-order acoustic modes can also be observed in hot $\delta$ Sct stars. Nevertheless, as shown in Fig. 1 this would imply that $\delta$ Sct pulsation covers the region between 205 and 870 $\mu$Hz {\it continuously}. According to current theory, the opacity mechanism acting in the HeII ionization zone, cannot excite modes spanning 16 radial orders for a star with parameters like those of HD 187547$^{28}$. Further support for the discovery of solar-like oscillations comes from spectroscopic observations that also indicate the presence of convective motions in the atmospheres of A and Am stars$^{29}$. In addition, signatures of granulation noise in $\delta$ Sct stars have been reported from photometric measurements$^{3}$. Opacity-driven pulsations are also observed in more massive stars (8-16 M$_{\odot}$), known as $\beta$ Cephei stars (in this case the opacity-mechanism acts in the ionization region of the iron-group elements). The unexpected detection of solar-like oscillations in such a star$^{30}$ (with a mass of 10 M$_{\odot}$), suggests that both types of pulsations, opacity-driven and stochastically excited, can coexist and can have overlapping frequency domains. The similar timescales of the different oscillation types imply a possible interaction between the two mechanisms.

\section*{References main article}

1.  Breger, M.  $\delta$ Scuti stars (Review). {\it Delta Scuti and Related Stars, ASP Conference Series} {\bf 210}, 3--42 (2000)\par

2.  Chaplin et al. Ensemble Asteroseismology of Solar-Type Stars with the NASA Kepler Mission. {\it Sci.} {\bf 332}, 213--216 (2011)\par

3.  Kallinger, T. \& Matthews, J.M. Evidence for Granulation in Early A-Type Stars. {\it Astrophys. J.} {\bf 711}, L35--L39 (2010)	\par

4.  Houdek, G., Balmforth, N.J., Christensen-Dalsgaard, J., Gough, D.O. Amplitudes of stochastically excited oscillations in main-sequence stars. {\it Astron. Astrophys.} {\bf 351}, 582--596, (1999) \par

5.  Samadi, R., Goupil, M.-J., Houdek, G. Solar-like oscillations in delta Scuti stars. {\it Astron. Astrophys.} {\bf 395}, 563--571 (2002)\par

6.  Koch, D.G. et al. Kepler Mission Design, Realized Photometric Performance, and Early Science. {\it Astrophys. J.}, {\bf 713}, L79--L86 (2010)\par

7.  Preston, G.W. The chemically peculiar stars of the upper main sequence. {\it ARA\&A} {\bf 12}, 257--277 (1974) \par

8.  Asplund, M., Grevesse, N., Sauval, A.J. Scott, P. The Chemical Composition of the Sun. {\it ARA \& A } {\bf 47}, 481--522 (2009) \par

9.  Carquillat,  J.-M., Prieur, J.-L. Contribution to the search for binaries among Am stars - VIII. New spectroscopic orbits of eight systems and statistical study of a sample of 91 Am stars. {\it Mon. Not. R. Astron. Soc.} {\bf 380}, 1064--1078 (2007)\par

10. Turcotte, S., Richer, J., Michaud, G., Christensen-Dalsgaard, J. The effect of diffusion on pulsations of stars on the upper main sequence - $\delta$ Scuti and metallic A stars. {\it Astron. Astrophys.} {\bf 360}, 603--616 (2000)\par

11. Balona, L. et al. Kepler observations of Am stars. {\it Mon. Not. R. Astron. Soc.} {\bf 414}, 792--800 (2011)\par

12.  Handler, G. Confirmation of simultaneous p and g mode excitation in HD 8801 and
$\gamma$ Peg from time-resolved multicolour photometry of six candidate 'hybrid'
pulsators. {\it Mon. Not. R. Astron. Soc.} {\bf 398}, 1339--1351(2009).\par

13.  Kurtz, D.W. Rapidly oscillating AP stars. {\it Mon. Not. R. Astron. Soc.} {\bf 200}, 807--859 (1982)\par

14.  Auriere M. et al., No detection of large-scale magnetic fields at the surfaces of Am and HgMn stars. {\it Astron. Astrophys.} {\bf 523}, A40 (2010)\par

15.  Stello, D., Chaplin, W.J., Basu, S., Elsworth, Y., Bedding, T.R. The relation between $\Delta \nu$ and $\nu_{\rm max}$ for solar-like oscillations. {\it Mon. Not. R. Astron. Soc.} {\bf 400L}, L80--L84 (2009)\par

16.   Houdek, G. Solar-type Variables. {\it Stellar Pulsation: Challanges for Theory and Observation, AIP Conference Proceedings} {\bf 1170}, 519--530 (2009)

17.  Samadi, R., Belkacem, K., Goupil, M.-J., Kupka, F., Dupret, M.-A. Solar-like oscillation amplitudes and line-widths as a probe for turbulent convection in stars. {\it IAUS} {\bf 239}, 349--357 (2007)\par

18.  Kjeldsen, H. et al. The Amplitude of Solar Oscillations Using Stellar Techniques, {\it Astrophys. J.} {\bf 682}, 1370--1375 (2008)\par

19. Ballot, J., Barban, C.  van't  Veer-Menneret, C. {\it Astron. Astrophys.}, submitted (arXiv:1105.4557v1)\par

20.  Kjeldsen, H. \& Bedding, T.R. Amplitudes of solar-like oscillations: a new scaling relation. {\it Astron. Astrophys.} {\bf 529}, L8 (2011) \par

21. Michel, E. et al. Intrinsic photometric characterisation of stellar oscillations and granulation. {\it Astron.   Astrophys.} {\bf 495}, 979--987 (2009) \par

22.  Gruberbauer, M., Kallinger, T., Weiss, W.W., Guenther, D.B. On the detection of Lorentzian profiles in a power spectrum: a Bayesian approach using ignorance priors. {\it Astron. Astrophys.} {\bf 506}, 1043--1053  (2009)\par

23.  Chaplin, W.J., Houdek, G., Karoff, C., Elsworth, Y., New, Mode lifetimes of stellar oscillations. Implications for asteroseismology. {\it Astron. Astrophys.} {\bf 500}, L21--L24 (2009)\par

24.  Baudin, F. et al. Amplitudes and lifetimes of solar-like oscillations observed by CoRoT. Red-giant versus main-sequence stars. {it  Astron. Astrophys.} {\bf 529}, A84 (2011)\par

25.  Samadi, R., Ludwig, H.-G., Belkacem, K., Goupil, M. J., Dupret, M.-A. The CoRoT target HD 49933. I. Effect of the metal abundance on the mode excitation rates. {\it Astron. Astrophys.} {\bf 509}, A15 (2010)

26. Chaplin, W.J. et al. On model predictions of the power spectral density of radial solar p modes. {\it Mon. Not. R. Astron. Soc.} {\bf 360}, 859--868 (2005)\par

27.  Chang, H.-Y. \& Gough, D.O. On the Power Distribution of Solar P MODES. {\it SoPh} {\bf 181}, 251--263 (1998)\par

28.  Pamyatnykh, A.A. Pulsational Instability Domain of $\delta$ Scuti Variables. {\it ASPC} {\bf 210}, 215--246 (2000)\par

29.  Landstreet, J.D. et al. Atmospheric velocity fields in tepid main sequence stars. {\it Astron. Astrophys.} {\bf 503}, 973--984 (2009)\par

30. Belkacem, K. et al. Solar-Like Oscillations in a Massive Star. {\it Sci} {\bf 324}  1540--1542 (2009)\par

{\bf Supplementary Information} is linked to the online version of the paper at www.nature.com/nature.

\section*{Acknowledgments}
Funding for this Discovery mission is provided by NASA's Science Mission Directorate. The authors wish to thank the entire {\it Kepler} team, without whom these results would not be possible. V.A., G.H. and G.Ho. were supported by the Austrian Fonds zur F\"orderung der wissenschaftlichen Forschung. V.A. also wants to thank Luca Fossati for his help. A.M. acknowledges the funding of AstroMadrid, who was also supported by Spanish grants. T.B. and D.S. acknowledge support from the Australian Reseach Council. T.L was supported by the Austrian Agency for International Cooperation in Education and Research. KU acknowledges financial support by the Deutsche Forschungsgemeinschaft.\par

\section*{Author contribution}
V.A. discovered the star among the {\it Kepler } targets, analysed it and found the solar-like oscillations (as a part of her PhD thesis), did spectroscopic analyses, frequency analyses, the test on the stochastic nature of the signal, interpretations and wrote the paper. G.H. had the idea for this project and supervised  V.A., helped with analyses, interpretations and writing the paper. T.L.C. contributed to the analyses of the stochastic modes and also to the test on the stochastic nature of the signal. A.O.T. observed the target spectroscopically at NOT, identified the star as an Am star and did spectroscopic analyses. A.M. contributed to the statistical test on the nature of the stochastic signal. T.K. helped interpretations, data analyses, writing the paper and delivered the mode lifetimes.  D.S. helped with data analyses, writing the paper and did Fig. 2. A.G. helped with theoretical interpretations and the time-fourier analyses. T.B. helped with interpretations and writing the paper. H.K. contributed to the analyses, also by supervising V.A. and is member of the KAI Steering Committee (Kepler Asteroseismic Investigation Steering Committee). J.C-D. helped with the theoretical support, writing the paper and is member of the KAI Steering Committee (Kepler Asteroseismic Investigation Steering Committee). T.L. confirmed the Am identification, excluding the Ap character of the star. G.C., A.F., and A.K. did spectroscopic analyses. P.DC. was PI and observer for the spectroscopic data from McDonald observatory. K.U. was CoI of the McDonald data and was coordinating the ground-based observations. H.B. was PI for the NARVAL spectrum and did spectroscopic analyses. G.Ho. and P.L. helped with theoretical interpretations and writing the paper. D.W.K. helped with the Am classification, writing the paper and is leader of the delta Scuti working group of the KASC (Kepler Asteroseismic Science Consortium). J.VC., C.A. and B.D.C. are part of the Kepler team and were involved in designing and operating the satellite. All co-authors contributed to discussions and commented on the manuscript.

\section*{Author information}
Reprints and permissions information is available at www.nature.com/reprints. Correspondence and requests for materials should be addressed to Victoria Antoci. E-mail: antoci@astro.univie.ac.at\par

 \begin{figure}

   \includegraphics[width=15cm, bb=77 479 540 796]{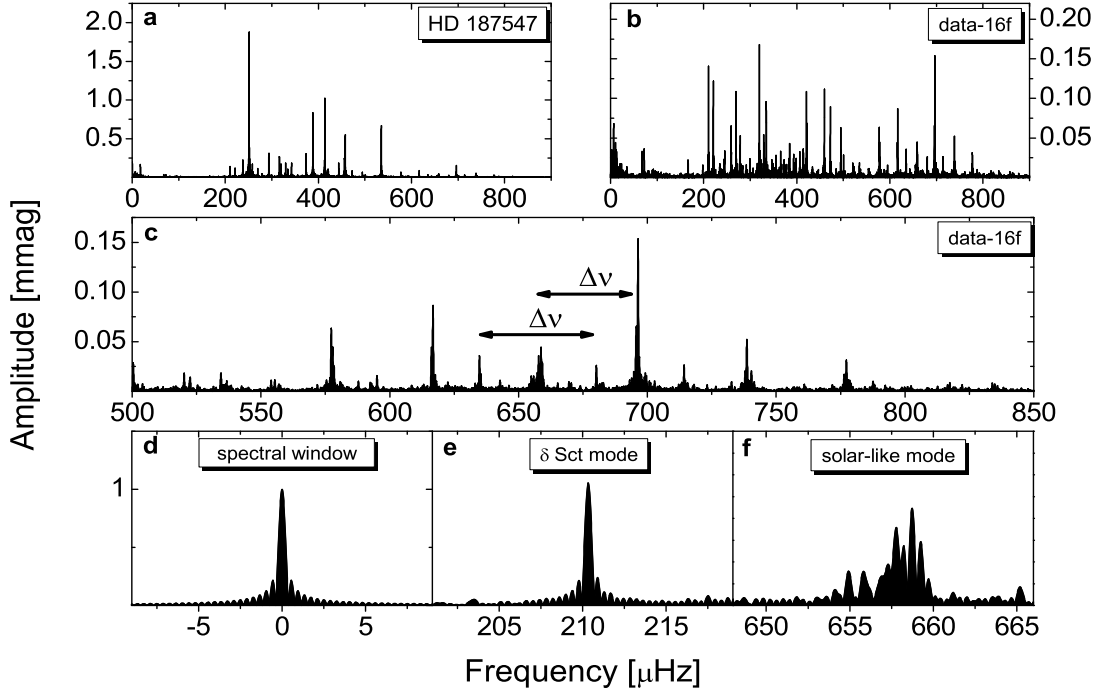}

  \caption{{\bf Fourier amplitude spectra of the {\it Kepler} light curve of HD 187547.}
{\bf a}, Fourier spectrum covering the entire frequency range in which significant signals were observed with a dominant frequency at 251 $\mu$Hz and an amplitude of 2 mmag, typical for a $\delta$ Sct star. {\bf b}, The multi-mode oscillations of HD 187547 are shown by subtracting 16 sinusoids corresponding to the most prominent oscillations, revealing a large number of additional significant frequencies.  {\bf c} The region between 500 and 850 $\mu$Hz shows a clear pattern of roughly equally spaced peaks, which we interpret as high-order consecutive radial overtones. The comb-like structure expected for high-order radial overtones is clearly visible. The broadened peaks suggest damped/re-excited solar-like oscillations. The black arrows denoted $\Delta \nu$ indicate the large separation between consecutive radial and dipole modes. {\bf d}, Spectral window. The shape of the window function is defined by the length and sampling of the data set. Any coherent signal will have the same profile. {\bf e}, Example for one of the modes driven by the opacity mechanism in HD 187547. {\bf f}, A supposed solar-like oscillation mode observed in HD 187547,  displaying a broadened structure suggestive of a short mode lifetime. }
   \end{figure}

\begin{figure}

   \includegraphics[width=8cm, bb= 146 291 420 588]{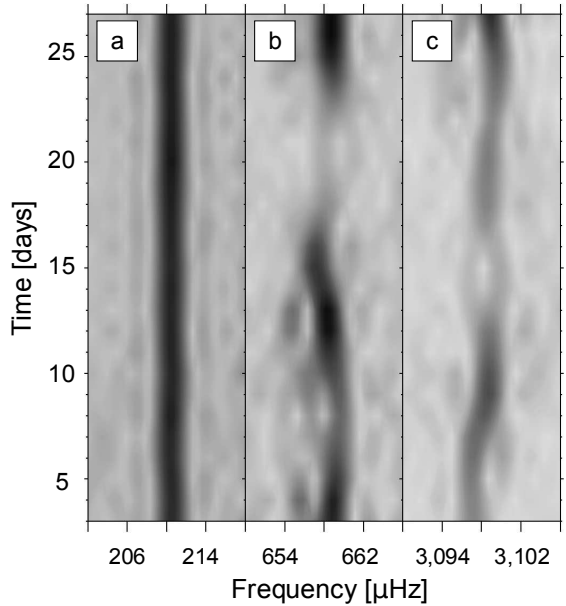}

  \caption{{\bf Time-Fourier spectrum.}
Here we again highlight the difference in temporal variability between the modes interpreted as stochastic modes and the coherent, opacity-driven peaks at low frequencies. The time-fourier spectrum was computed with a running filter of full-width at half-maximum = 5 days, comparable to the mean mode lifetime. {\bf a}, An opacity-driven mode (the same as displayed in Fig. 1e) showing temporal stability in the $\delta$ Sct frequency region. {\bf b}, Stochastic mode observed in HD 187547, displaying an erratic behaviour as expected for solar-like oscillations (the same as displayed in Fig. 1f). {\bf c}, For comparison, a stochastic oscillation mode observed in the Sun. The solar data were obtained from the SOHO VIRGO instrument. The data set has the same length and sampling as for HD 187547, that is, 30 days and 1 min, respectively. Further details of frequency analyses and tests on artificial data sets (Supplementary Fig. 1) to verify our interpretation are in the Supplementary Information. }
   \end{figure}

\begin{figure}

   \includegraphics[width=15cm]{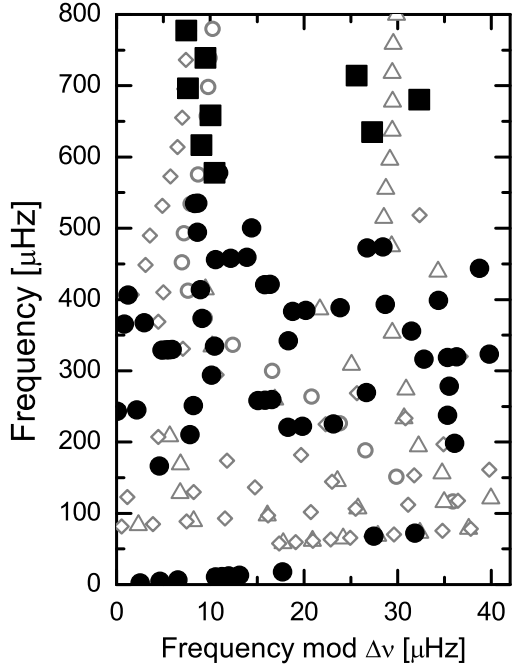}

  \caption{{bf \'{E}chelle diagram of HD 187547.}
Here we plot 69 extracted frequencies as a function of frequency modulo the large separation ($\Delta \nu$ = 40.5~$\mu$Hz). Frequencies equally spaced in $\Delta \nu$ will form vertical ridges in the  \'{e}chelle diagram. To guide the eye, we show theoretically predicted frequencies of pulsation modes for a 1.85~M$_{\odot}$ stellar model. Ridges of $l=0$ modes are represented by open circles, $l=1$  by open triangles and $l=2$  by open diamonds. Detailed modeling of the star is beyond the scope of this paper. The supposed solar-like modes (filled squares) between 500 and 870 $\mu$Hz show clear ridges, as expected for high-order acoustic oscillations similar to what is observed in solar-like stars. The lower frequencies that we attribute to $\delta$ Sct pulsation (filled circles) excited by the opacity mechanism show no obvious regular patterns.}
   \end{figure}

\end{document}